\documentclass{article}
\usepackage{arxiv}
\usepackage[utf8]{inputenc} % allow utf-8 input
\usepackage[T1]{fontenc}    % use 8-bit T1 fonts
\usepackage{url}            % simple URL typesetting
\usepackage{booktabs}       % professional-quality tables
\usepackage{amsfonts}       % blackboard math symbols
\usepackage{amsmath}
\usepackage{amssymb}
\usepackage{siunitx}
\usepackage{nicefrac}       % compact symbols for 1/2, etc.
\usepackage{microtype}      % microtypography
\usepackage{lipsum}		% Can be removed after putting your text content
\usepackage{doi}
\usepackage[numbers]{natbib}
\usepackage[normalem]{ulem}
\usepackage{graphicx}
\usepackage{mathtools}
\usepackage{xcolor} 
\usepackage{wasysym}
\usepackage{cuted}
\usepackage{hyphenat}
\usepackage{tikz}
\usepackage{textcomp}
\usepackage{bookmark}

\newcommand{\Figref}[1]{Fig.~\ref{#1}}
\newcommand{\Figrefa}[1]{Fig.~\ref{#1}(a)}
\newcommand{\Figrefb}[1]{Fig.~\ref{#1}(b)}
\newcommand{\Figrefc}[1]{Fig.~\ref{#1}(c)}
\newcommand{\Figrefd}[1]{Fig.~\ref{#1}(d)}
\newcommand{\Figrefe}[1]{Fig.~\ref{#1}(e)}
\newcommand{\Figreff}[1]{Fig.~\ref{#1}(f)}

\newcommand{\eqnref}[1]{Eq.~\ref{#1}}

\newcommand\copyrighttext{%
  \footnotesize \textcopyright 2024 IEEE. Personal use of this material is permitted.
  Permission from IEEE must be obtained for all other uses, in any current or future 
  media, including reprinting/republishing this material for advertising or promotional 
  purposes, creating new collective works, for resale or redistribution to servers or 
  lists, or reuse of any copyrighted component of this work in other works.  DOI: \href{https://doi.org/10.1109/TTHZ.2024.3388254}{10.1109/TTHZ.2024.3388254}}
\newcommand\copyrightnotice{%
\begin{tikzpicture}[remember picture,overlay]
\node[anchor=south,yshift=10pt] at (current page.south) {\fbox{\parbox{\dimexpr\textwidth-\fboxsep-\fboxrule\relax}{\copyrighttext}}};
\end{tikzpicture}%
}

\begin{document}
% *** IEEE Copyright notice with TikZ ***
% 
\copyrightnotice
\title{Modeling of antenna-coupled Si MOSFETs in the Terahertz Frequency Range}
\renewcommand{\shorttitle}{Modeling of antenna-coupled Si MOSFETs in the Terahertz Frequency Range}
\hypersetup{
pdftitle={Modeling of antenna-coupled Si MOSFETs in the Terahertz Frequency Range},
pdfsubject={physics.app-ph, physics.cond-mat},
pdfauthor={Florian~Ludwig, Jakob~Holstein, Anastasiya~Krysl, Alvydas~Lisauskas, Hartmut~G.~Roskos},
pdfkeywords={terahertz, detection, MOSFET, TSMC, hydrodynamic, Gaussian beam, power coupling, antenna simulation},
}

\author{Florian~Ludwig, Jakob~Holstein, Anastasiya~Krysl, Alvydas~Lisauskas, Hartmut~G.~Roskos % <-this % stops a space
\thanks{F. Ludwig, A. Krysl, J. Holstein and H. G. Roskos are with Physikalisches Institut, Johann Wolfgang Goethe-Universität, DE-60438 Frankfurt am Main, Germany (e-mail: roskos@physik.uni-frankfurt.de)}% <-this % stops a space
\thanks{A. Lisauskas is with (i) Institute of Applied Electrodynamics and Telecommunications, Vilnius University, LT-10257 Vilnius, Lithuania, and (ii) Physikalisches Institut, Johann Wolfgang Goethe-Universität, DE-60438 Frankfurt am Main, Germany}}%

% \today 

\maketitle

\begin{abstract}
\textcolor{black}{We report on the modeling and experimental characterization of Si CMOS detectors of terahertz radiation based on antenna-coupled field-effect transistors (TeraFETs). The detectors are manufactured using TSMC's 65-nm technology. We apply two models -- the TSMC RF foundry model and our own ADS-HDM  -- to simulate the Si CMOS TeraFET performance and compare their predictions with respective experimental data. Both models are implemented in the commercial circuit simulation software Keysight Advanced Design System (ADS). We find that the compact model TSMC RF is capable to predict the detector responsivity and its dependence on frequency and gate voltage with good accuracy up to the highest frequency of 1.2~THz covered in this study. This frequency is well beyond the tool's intended operation range for 5G communications and 110-GHz millimeter wave applications. We demonstrate that our self-developed physics-based ADS-HDM tool, which relies on an extended one-dimensional hydrodynamic transport model and can be adapted readily to other material technologies, has high predictive qualities comparable to those of the foundry model. We use the ADS-HDM to discuss the contribution of diffusive and plasmonic effects to the THz response of Si CMOS TeraFETs, finding that these effects, while becoming more significant with rising frequency, are never dominant. Finally, we estimate that the electrical NEP (perfect power coupling conditions) is on the order of 5~pW/$\sqrt{\rm{Hz}}$ at room-temperature. }
\end{abstract}


\section{Introduction}
Antenna-coupled field-effect transistors (TeraFETs) have emerged as powerful devices for a variety of terahertz (THz) applications such as frequency multiplication and mixing \cite{Dyakonov1996, Glaab2010, Giliberti2013,LisauskasSubharmonic,Yuan3DFourier, Wiecha2021}, as well as direct THz power detection \cite{Dyakonov1996, Knap2009, Ojefors2009, Lisauskas2009, Popov2011,Dyer2012a,Sun2012,Boppel2012,Nagatsuma2013,Knap2013,BauerAntennaCoupled,GrzybTHzDirectDetector,Sun2020,Andree2022}. The detectors can be readily integrated in electronic circuits and provide sufficient sensitivity at room temperature which makes them well suited for applications. Among these, TeraFETs fabricated in Si CMOS technology have reached impressive performance levels \cite{Boppel2012,IkamasBroadbandTerahertz} with a high sensitivity over almost the entire THz frequency range. \textcolor{black}{With the growing variety of possible applications of field-effect transistors (FETs) in the THz range, the need for reliable design and modeling platforms for such devices has increased significantly.} Here, we focus on detector simulation and experimentally validate our results. In \cite{Ludwig2019} we presented our simulation approach for TeraFETs -- acronym: ADS-HDM. The ADS-HDM is implemented into the circuit simulation environment Keysight ADS - an industry standard. Its implementation is based on an extended version of the one-dimensional (1D) hydrodynamic transport model (HDM), which has been used in \cite{Dyakonov1996}. The frequency-dependent channel impedance is determined from the self-consistent solution of the HDM equations. \textcolor{black}{This physics-based model accounts for the mixing process in the FET channel involving the charge-carrier density waves (plasma waves). They are excited by the incident THz radiation via an antenna element which couples the radiation asymmetrically to the FET channel, either via the source-gate or the drain-gate port \cite{Dyakonov1993,Boppel2012,Drexler2012,Boppel2016,Bauer2019}.} The evidence for the existence of plasma waves (which are overdamped in the case of Si CMOS at room temperature) has been strengthened by experimental findings described in \cite{Drexler2012, Bandurin2018a, Soltani2020, Caridad2024}. In this work, we present and discuss two simulation approaches to simulate the detector performance of Si CMOS TeraFETs, our in-house tool ADS-HDM as well as the TSMC RF foundry model (TSMC RF) of the Taiwan semiconductor manufacturing company (TSMC). Prior to the simulations we propose an approach to determine the exact THz input power present at the transistor terminals when the detector is illuminated with a Gaussian intensity profile. 
The model predictions are compared quantitatively with the optical voltage responsivity determined between 0.4 and 1.2~THz and put into perspective with simulation results obtained with the TSMC RF. The ADS-HDM tool allows to switch the plasmonic effects on and off. Using this feature, we quantify the plasmonic contribution to the THz detector responsivity. Finally, we discuss the upper sensitivity limits of the 65-nm Si CMOS TeraFETs.

\section{Hydrodynamic Model description}
In order to simulate the excitation of density waves of the charge carriers in the channel of the Si MOSFET detectors, we use the following 1D hydrodynamic transport equations
\begin{subequations}\label{eqn:HDM}
  \begin{align}
     \partial_t n & = \frac{1}{q} \partial_x j \,, \label{eqn:Continuity} \\
     \partial_t j & = \frac{q^2 n}{m^*} E_x + \frac{q k_B T_L}{m^*} \partial_x n + \frac{1}{q} \partial_x \left(\frac{j^2}{n}\right) - \frac{j}{\tau_p} \,, \label{eqn:Momentum Balance}
  \end{align}
\end{subequations}
consisting of the continuity equation \eqnref{eqn:Continuity} and the momentum balance equation \eqnref{eqn:Momentum Balance}, both written in terms of the current density $j = - qnv$, where $q$ is the electron charge, $n=n(x)$ represents the carrier sheet density in the FET's inversion layer along the channel ($x$-direction), and $v=v(x)$ stands for the local mean carrier velocity. Note that -- contrary to the model presented in \cite{Dyakonov1996,Dyakonov1993} --, the second term on the right hand side of \eqnref{eqn:Momentum Balance} also takes carrier diffusion into account. Other quantities of the equations are the longitudinal electric field $E_x = - \partial_x \varphi$  as determined by the channel potential $\varphi (x)$, the electron effective mass ($m^* = 0.26\, m_0$ for electrons in Si in units of the free-electron mass $m_0$), and the momentum relaxation time $\tau_p$ of the electrons. The temperature of the crystal lattice is assumed to be room temperature ($T_L$ = 293 K). $k_B$ denotes the Boltzmann constant.  The dependence of $n(x)$ on the externally applied gate voltage is taken into account by the unified charge control model (UCCM) for n-channel MOSFETs \cite{Shur1992}
\begin{align} \label{eqn:UCCM}
n(x) = \frac{C_{ox} \gamma \, U_{th}}{q} \cdot \ln{\left(1+\frac{1}{2}\exp{\left(\frac{ U_{gc}(x) - U_T }{\gamma U_{th}}\right)}\right)} \,,
\end{align}
where $C_{ox} = \frac{\epsilon_R \epsilon_0}{d}$ is the channel capacitance per unit area for a thickness $d$ of the gate insulator. For $d$, we assume a value of 3.05~nm.\footnote{Note that the value of $d$ represents an effective insulator thickness. $d = d_{phys} + d_{qc}$ is the sum of the physical insulator thickness ($d_{phys} = 2.6~\rm{nm}$) and a quantum correction of $d_{gc} = 0.45~\rm{nm}$ arising from the quantum-confinement effect of electrons at the Si-to-SiO\textsubscript{2} interface in nanoscale MOSFETs \cite{Saad2010}. The value of $d_{phys} = 2.6~\rm{nm}$ is taken from the information sheet of the 65-nm CMOS foundry process.}
\begin{figure}[!t]
\centering
\includegraphics[width=3.5in]{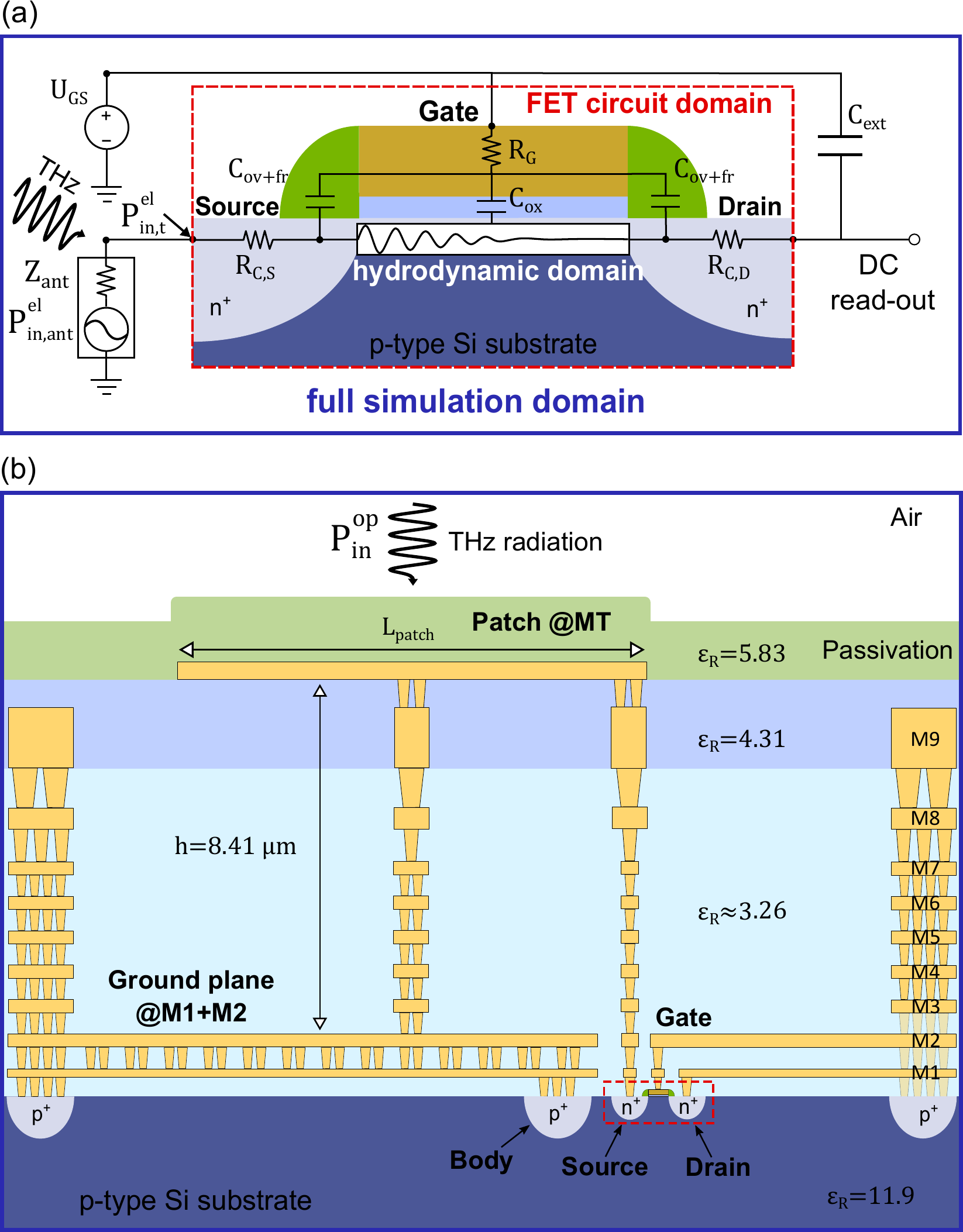}
\caption{(a) Schematic of the ADS-HDM simulation tool: The blue box (full simulation domain) represents the THz detector model (is taken in the same manner for the TSMC RF foundry model), here for source radiation coupling conditions. At its core (hydrodynamic domain), the hydrodynamic transport equations are used to calculate the rectified signal of the detector on the basis of plasma-wave-modified resistive mixing in the transistor's channel. The equations are solved in the Keysight ADS simulator itself. The FET parasitics are taken into account in the FET circuit domain. 
Parasitic components considered are the source and drain overlap $C_{ov}$ and fringe capacitances $C_{fr}$, the contact resistances $R_{C,S}$ and $R_{C,D}$, and the gate resistance $R_{G}$. The antenna parameters including the antenna impedance $Z_{ant}$ are estimated from EM wave simulations (CST Studio Suite). For details of the HDM implementation, we refer to \cite{Ludwig2019}. \\ (b) Cross-sectional view of the detector implementation in TSMC 65-nm process. The patch antenna is implemented underneath a passivation layer in the top metal layer MT. The ground plane is placed in lowest to metal layers M1 and M2.}
\label{fig:MOSFET_HDM}
\end{figure}
$\epsilon_R = 3.9$ is the relative permittivity of the gate insulator, and $\epsilon_0$ the vacuum permittivity. $U_T$ and $U_{th} = \frac{k_B T_L}{q}$ are the threshold voltage and the thermal voltage, respectively. $\gamma$ represents a non-ideality factor, which determines the subthreshold slope of the drain current with regards to the gate-to-source voltage $U_{GS}$.
The gate-to-channel voltage $U_{gc}(x) = U_{GS} - \varphi(x)$ is given by the difference of $U_{GS}$ and the channel potential $\varphi(x)$, the latter being determined self-consistently by jointly solving Eqs.~\ref{eqn:HDM}a, \ref{eqn:HDM}b and \ref{eqn:UCCM}. \\
The numerical evaluation of this set of equations is performed in the circuit simulation environment of Keysight ADS using a \textcolor{black}{novel} distributed-element method approach to integrate the transport equations as presented in \cite{Ludwig2019} for AlGaN/GaN HEMTs. The steady-state solution  
is obtained by the harmonic balance technique. A schematic of the ADS-HDM implementation is shown in \Figrefa{fig:MOSFET_HDM}. 
A significant advantage of this simulation approach versus recently developed TCAD models for the THz frequency range \cite{Bhardwaj2014,Jungemann2016,Liu2019TCAD,Linn2020} is that in the case of circuit based models like ADS-HDM \cite{Ludwig2019} or recently developed compact SPICE models \cite{Liu2019Spice,Liu2022SPICE}, extrinsic and parasitic device components are readily included in the full simulation domain (as indicated in \Figrefa{fig:MOSFET_HDM}), while for TCAD simulations they have to be included via more complex mixed mode simulation methods \cite{Grasser2000}. 
At the same time, the full physical model for the description of the plasma-wave mixing process in the transistor's channel is retained at the model's core (hydrodynamic domain). By manipulating or switching off individual terms in Eqs.~\ref{eqn:HDM}a, \ref{eqn:HDM}b and \ref{eqn:UCCM}, one can study and identify the influences of the various physical processes included in the equations. \\
A number of input values for the parameters of the simulations, as presented in Table~\ref{tab:FETparameter}, can be obtained from the measured DC drain-to-source resistance (see below). This makes it easier to test whether variations of the THz performance of each TeraFET may be caused by fabrication-related issues which also affect the DC device properties.

\section{65-nm Si CMOS TeraFETs}
\begin{figure}[!t]
\centering
\includegraphics[width=3.5in]{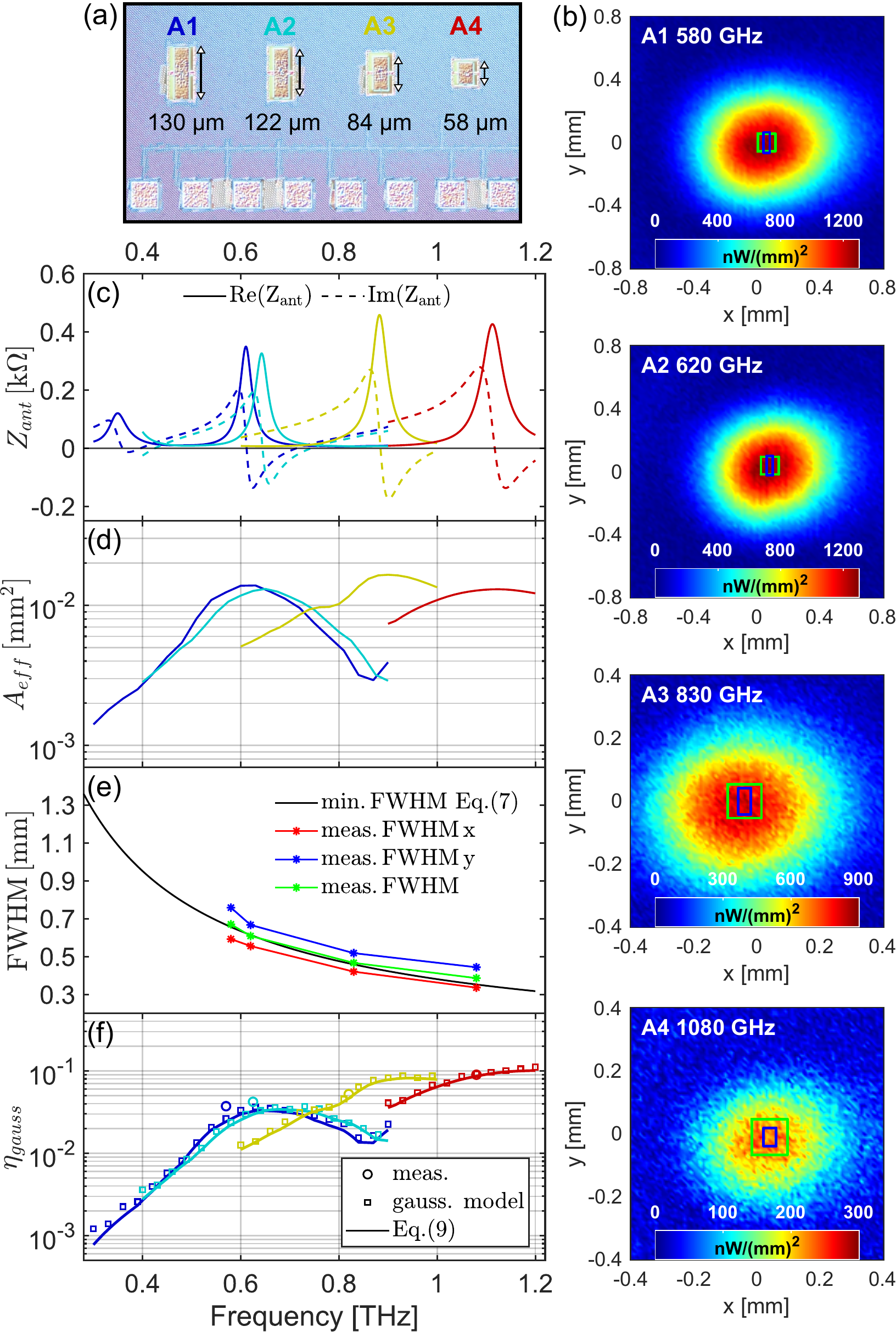}
\caption{(a) Top-view micrograph of the array of four detectors (A1-A4) showing the rectangular patch antennas; the corresponding patch lengths are listed. All patches have the same width of 40~$\mu$m and height of 8.41~$\mu$m above the ground-plane. (b) Direct measurement of the Gaussian beam intensity profile $I_{gauss}(x,y)$ obtained close to the resonant frequencies of the patch antennas A1-A4. The physical dimensions ($L_{patch} \cdot W_{patch}$) of each patch are indicated as blue rectangles, while the simulated effective area $A_{eff}$ is shown in green. (c) Simulated real (solid line) and imaginary part (dashed line) of the antenna impedance $Z_{ant}$ and (d) effective area $A_{eff}$ obtained from EM wave simulations (CST Studio Suite). (e) Extracted FWHM in $x$ (red), $y$ (blue) and estimated mean FWHM (green) of the Gaussian beam intensity profile measured in the focal point of a 2'' OAP mirror as presented in (b). The black solid line shows the calculated minimum FWHM from Eq.~(\ref{eqn:Df}) for $d_L$ = 30~mm and $f_L$ = 50.8~mm. (f) Antenna Gaussian beam
coupling efficiency $\eta_{gauss}$ for each detector (A1-A4) extracted from direct measurements of the Gaussian beam intensity profile (open dots). In addition we present $\eta_{gauss}$ for arbitrary frequencies assuming (i) an ideal Gaussian beam intensity profile (open squares) and (ii) solving Eq.~(\ref{eqn:effgauss}) numerically (solid lines). In both cases the FWHM of the Gaussian beam is determined from Eq.~(\ref{eqn:Df}) (as shown as black line in (e)).}
\label{fig:antenna_parameters}
\end{figure}

The detectors studied in the following are designed for a commercial 65-nm complementary metal-oxide-silicon (CMOS) foundry process (CLN65LP, type 1.2V\_mLow\_Vt\_MOS) of the Taiwan Semiconductor Manu\-fac\-tu\-ring Company (TSMC). \Figrefb{fig:MOSFET_HDM} displays a cross-sectional schematic view of the detector design. In brief, a narrow-band rectangular patch antenna, which is placed in the last metal layer M10, collects the THz radiation and guides the signal through a via at one of the narrow sides of the patch to the source of a single transistor placed directly under the edge of the patch. The height of the patch above the ground plane (metal layer M1 and M2) is $8.41~\rm{\mu}$m. \\
In the following, four different devices are studied with respect to their detector performance. The devices differ only in terms of their respective patch length $L_{patch}$ (see micrograph with the values in \Figrefa{fig:antenna_parameters}), which results in different resonant frequencies. The patch width is fixed to 40~$\rm{\mu}$m. A 1.3-$\rm{\mu}$m-thick passivation layer covers the devices. An important figure of merit, which determines the efficiency of our detectors, is the optical voltage responsivity $\mathcal{R}_{V}^{op}$.\footnote{In this work, we aim to model explicitly the optical responsivity of the detectors. One distinguishes between \textit{optical} and \textit{cross-sectional} responsivity and NEP \cite{Bauer2019}. The cross-sectional responsivity is employed if the antenna cross-section is smaller than the beam size. It is calculated as the optical responsivity multiplied by the ratio of the (measured or calculated) beam cross-sectional area at the antenna to the (calculated) antenna cross-sectional area \cite{Javadi2021}.}
It is defined as the rectified voltage (or current) signal between the drain and source terminal $\Delta U_{DS}$ with regards to the available optical THz input power $P_{in}^{op}$, the total power of the incident THz beam in free-space \cite{Javadi2021,Ferreras2021}:
\begin{align}
    {\mathcal{R}_{V}^{op}} = \frac{\Delta U_{DS}}{P_{in}^{op}}.
    \label{eqn:voltage_responsivity}
\end{align}
In this work, the responsivity is measured with the help of an optoelectronic spectroscopy system (Tera\-Scan 1550, Toptica Photonics AG, Graefelfing (Munich)), where the optoelectronic detector is replaced by our TeraFETs. $\Delta U_{DS}$ is measured using the lock-in technique. For this purpose, the bias to the photoconductive emitter is modulated with a sine wave at $f_{mod}$ = 7.62~kHz. \textcolor{black}{A schematic drawing of the measurement setup is displayed in \Figrefa{fig:LIA_correction}.} Continuous-wave THz radiation between 0.1 and 1.2~THz is generated by an InGaAs photomixer emitter. Its radiation is guided to the Si CMOS TeraFET via four off-axis paraboloid (OAP) mirrors with 90°-beam-deflection geometry. The THz beam path is purged by dry nitrogen in order to minimize radiation absorption by water vapor. 
The total power $P_{in}^{op}$ of the incoming THz beam in front of the Si CMOS TeraFET, placed in the focal point of the last OAP mirror, is measured using a calibrated large-area Golay cell. 

The peak-to-peak rectified voltage between drain and source contacts is obtained as $\Delta U_{DS} = \alpha_{LIA} \cdot U_{LIA}$, where $U_{LIA}$ is the lock-in amplifier's output signal. It is \textcolor{black}{proportional} to the RMS of the first Fourier fundamental component of the detector signal. $\alpha_{LIA}$ represents a lock-in correction factor. % in order to relate the detector calibration measurements, which are sensitive to the full-waveform peak-to-peak response, . T
For an ideal sinusoidal modulation (e.g by electronic chopping) of an emitter, the lock-in correction factor is found to be a constant value of $\alpha_{LIA,corr}=2\sqrt{2}$. \textcolor{black}{ However, inspection with an oscilloscope shows that the THz wave is not modulated sinusoidally, but rather exhibits a more square-wave-like waveform with a detailed shape which depends on the beat-note frequency of the laser radiation. % investigated the influence of electronic chopping a InGaAs photomixer at different THz frequencies and found $
Measured results and the derivation of the THz-frequency-dependent values of % A detailed analysis of the waveforms of the measurement setup and the estimation of 
$\alpha_{LIA,corr}$ are discussed in Appendix A}.
$\alpha_{LIA,corr}$ is found to increase linearly with frequency from $\sim2.4$ (at 0.1~THz) to $\sim3.1$ (at 1.2~THz). 

\section{Power coupling theory}
For a proper comparison of the simulated optical voltage responsivity with the measured one to be addressed later, the electrical THz input power $P_{in,t}^{el}$ at the source transistor terminal (here for source coupling conditions) has to be estimated. For this purpose, we need to consider multiple losses suffered by the incoming radiation before it is converted into the electrical input signal. % of the THz optical input power 
$P_{in}^{op}$ denotes the power of the incoming THz beam directly before %present at 
the passivation layer. 
\textcolor{black}{The electric power $P_{in,ant}^{el}$, which one obtains from this optical power at the antenna} % This optical power is On its way to the antenna and from there to the source electrode (see \Figrefb{fig:MOSFET_HDM}) only a part of this power is %interface when converted into the electrical signal. The 
%  First, we have to determine the electrical THz input power $P_{in,ant}^{el}$ originating from the antenna after all optical and dissipative losses, which can be obtained from
is calculated in the following way:
\begin{align}
    P_{in,ant}^{el} = \eta_{op} \cdot \eta_{scat} \cdot \eta_{gauss}(\nu) \cdot P_{in}^{op} \,.
    \label{eqn:electrical_power}
\end{align}
\textcolor{black}{Here, $\eta_{op}$ takes into account all possible optical losses of the incident THz radiation on the path from the source to the detector, which are not considered by the electromagnetic (EM) wave simulations. For example absorption and reflection losses associated with the substrate and a lens, which may be attached to it \cite{Bauer2019,Ferreras2021}.
%Here, $\eta_{op}$ takes account of beam attenuation on the way to the antenna that is not considered in the antenna simulations itself,  incurred %ical losses represents the optical path coupling efficiency, which accounts for of the incident THz radiation associated with %due to Drude 
%by absorption and reflection losses related to the substrate and a lens, which may be attached to it \cite{Bauer2019,Ferreras2021}.} %, and due to reflections at interfaces (where the materials' refractive indices differ, $n_1\neq n_2$).} %Such losses can occur, for example, if superstrate lenses are used, or -- in case of backside illumination -- in the substratein conjunction with a dielectric lenses \cite{Bauer2019,Ferreras2021}. 
In this work, we set $\eta_{op} = 1$, since %in our case backside illumination is obstructed by the ground plane.} Instead, 
the incident THz light is coupled directly from free space to the antenna as indicated in \Figrefb{fig:MOSFET_HDM}, neither passing through a substrate nor a substrate/superstrate lens (backside illumination is obstructed by the ground plane).} \textcolor{black}{$\eta_{scat}$ represents  
%incorporates residual scattering of incident power in an receiving antenna element, also stated as
the absorption efficiency of the antenna. It %does not include mode mismatch (which is taken care of by the next factor, $\eta_{gauss} (\nu)$),  
 includes the residual scattering losses that occur in a receiving antenna element even in case of a perfectly matched load. For large planar antennas it is known to be $\eta_{scat} \approx 0.5$ \cite{Andersen2005,Javadi2021}. The final loss factor is the frequency-dependent Gaussian beam coupling efficiency $\eta_{gauss} (\nu)$. From a geometrical point of view, $\eta_{gauss} (\nu)$ considers collection losses resulting from the discrepancy between the spatial extension %dimensions 
 of the incident Gaussian beam's intensity profile, given by \cite{Yariv2007}
\begin{align}
\label{eqn:Igauss}
  I_{gauss}(r,\varphi) = I_0 \cdot \exp\left(-\frac{k_0}{z_R} \cdot r^2\right) \,, 
\end{align}
and the effective area of the antenna} \cite{Volakis2007}
\begin{align}
\label{eqn:Aeff}
A_{eff} & = \frac{{\lambda_0}^2}{4\pi} \cdot G = \frac{{\lambda_0}^2}{4\pi} \cdot \epsilon_{ant} \cdot D \, .
\end{align}
Here $z_R=\frac{\pi\nu}{4 c_0} \cdot {D_f}^2$ is the Rayleigh range and $k_0=\frac{2\pi}{\lambda_0}=\frac{2\pi\nu}{c_0}$ the wave vector in free-space with speed of light in vacuum $c_0$. $I_0$ denotes the intensity in the center of the Gaussian beam. $D_{f}$ represents the minimum $1/e^2$ Gaussian beam diameter
\begin{align}
\label{eqn:Df}
D_{f} & = \frac{\sqrt{2} \cdot \rm{FWHM}}{\sqrt{\rm{ln}(2)}} = \frac{4 \cdot f_L \cdot c_0}{d_L \cdot \pi \cdot \nu}\, ,
\end{align}
where $d_L$ is the diameter of the collimated optical beam before the focusing element and $f_L$ is the focal length of the focusing element (in this work an OAP mirror with $f_L$ = 2" = 50.8 mm).
The total incident THz power is distributed over $I_{gauss}(r,\varphi)$ such that
\begin{align}
\begin{split}
P_{gauss} = P_{in}^{op} &= \int_{A} I_{gauss}(r,\varphi) \, \rm{d}A \, \\
          &= \int_{0}^{\infty} \int_{0}^{2\pi} r \cdot  I_{gauss}(r,\varphi) \,\rm{d}\varphi \rm{d}r \\
          &= 2\pi \cdot [1 - \exp{(-\alpha r^2})]_0^\infty = \frac{I_0\pi}{\alpha} \,
          \label{eqn:power}
\end{split}
\end{align}
with $\alpha=\frac{k_0}{z_R}=\frac{8}{{D_f}^2} = \frac{4 \cdot \rm{ln}(2)}{\rm{FWHM}^2}$.
$A_{eff}$ of the receiving antenna element is determined in the direction of maximum gain $G$.\footnote{A rectangular patch excited in its fundamental mode has a maximum gain in the direction perpendicular to the antenna plane ($\Phi=0^\circ$). Based on our experimental setup (see Appendix A) we assume perfect perpendicular illumination of the optical beam to the antenna plane to estimate the effective area.} $\epsilon_{ant}$ is the antenna efficiency and $D$ the antenna directivity. Both, $A_{eff}$ and $\epsilon_{ant}$, can be obtained from EM simulations (CST Studio Suite, shown in \Figrefd{fig:antenna_parameters}) or from direct measurements (e.g. absolute gain methods or gain transfer methods). For more details about the different methods to estimate the effective area of a THz detector, we refer to \cite{Javadi2021}. \textcolor{black}{Our EM wave simulations take the full three-dimensional structure of the Si CMOS TeraFET into account (\textcolor{black}{i.e., all structures shown in} the cross-sectional view of \Figrefb{fig:MOSFET_HDM}). This includes the vias from the antenna to the transistor \textcolor{black}{and the ground-plane}, the surrounding metallization, the different dielectric layers and the passivation layer. The lumped element port is placed at the position of the transistor.}  \textcolor{black}{In general, the EM wave simulations of $\epsilon_{ant} (\nu)$ and thus also of $\eta_{gauss} (\nu)$ consider both conduction losses of the metallization and possible dielectric losses due to the surrounding dielectric layers. Consistent with the real device, the metallization is simulated as lossy copper, so conduction losses are considered in our simulations. Dielectric losses are not considered here because the dielectric layers are made of insulator materials with zero loss tangent.} \\
We want to clarify that the Gaussian beam coupling efficiency $\eta_{gauss} (\nu)$ is closely related but not identical to the \textit{gaussicity} \cite{Filipovic1993} - a quantity typically employed for substrate-lens-assisted antennas. These typically are configurations, where a dielectric lens (e.g. hyper-hemispherical silicon lens) is placed onto the detector's substrate to enhance the Gaussian beam coupling efficiency under backside illumination conditions \cite{VanRudd2002,Ikamas2018,Filipovic1993}. The gaussicity can be estimated from the overlap integral of the normalized simulated antenna pattern and the best focused Gaussian beam in the angular domain \cite{Filipovic1993, Ferreras2021}. This calculation imposes that the effective area of the antenna is larger than (or of the same size as) the Gaussian beam diameter at the position of the antenna plane (after the dielectric lens and the substrate). \\
In this work, we determine $\eta_{gauss} (\nu)$ for free-space optical coupling conditions, where the latter condition is not fulfilled. In the following we assume that the detector is positioned in the focal point of a focusing element such as an OAP mirror (see \Figrefa{fig:LIA_correction} in Appendix A) and that the evolution of the optical beam in the experiment can be treated on basis of Gaussian beam theory \cite{Yariv2007}. Under these circumstances $\eta_{gauss} (\nu)$ can be calculated for arbitrary frequencies from
\begin{align}
\label{eqn:effgauss}
\begin{split}
\eta_{gauss} (\nu) & = \frac{P_{ant}}{P_{gauss}} = 1 - \frac{4\beta (\nu)}{\pi} \approx \rm{ln}(2) \cdot \frac{A_{eff}(\nu)}{A_{spot}(\nu)}\\
\beta(\nu)         & = \int_{0}^{\frac{\pi}{4}} \exp\left(-\frac{\rm{ln}(2)\cdot A_{eff}(\nu)}{\rm{FWHM}(\nu)^2} \cdot \frac{1}{\cos^2(\varphi)}\right) \rm{d}\varphi  \, ,
\end{split}
\end{align}
where $\rm{FWHM}=\sqrt{\rm{ln}(2)/2} \cdot D_{f}$ is the frequency-dependent minimum full-width at half maximum of the incident Gaussian beam given in the antenna plane (in focal point of the OAP mirror) and $A_{spot} =  \pi \cdot (\rm{FWHM}/2)^2$ is the spot size of the beam. Note that the integral for $\beta(\nu)$ has no analytical solution and needs to be solved numerically. However, we find that $\beta$ can be approximated by  $\beta \approx \pi/4 -  \rm{ln}(2) \cdot A_{eff}/A_{spot}$, which leads to $\eta_{gauss} \approx \rm{ln}(2) \cdot A_{eff}/A_{spot}$. For details on the derivation of the equation set Eq.~(\ref{eqn:effgauss}), we refer to Appendix B. \\
\textcolor{black}{Precise knowledge of $d_L$ is therefore required to determine the Gaussian beam intensity profile at the detector position.} For this purpose, we \textcolor{black}{estimate} the FWHM (\textcolor{black}{or} $D_f$) at the focal point of the OAP mirror by direct measurements of the Gaussian beam intensity profile $I_{gauss}(x,y)$ using an automated X-Y micrometer translation stage as shown in \Figrefb{fig:antenna_parameters} at the respective \textcolor{black}{resonant frequency} of each detector A1-A4. \textcolor{black}{We want to emphasize that $I_{gauss}(x,y)$ may have to be de-embedded when using highly directional antennas with a narrow beamwidth. The 3-db beamwidth (angular width) of the patch antennas in this work is roughly $90$° over a wide range of frequencies.} From $I_{gauss}(x,y)$ we determine the FWHM of the Gaussian distribution alongside X (shown as red line) and Y (shown as green line) using a Gaussian fit model (see \Figrefe{fig:antenna_parameters}). \textcolor{black}{Then the} mean FWHM can be obtained from $\rm{FWHM}=\sqrt{\rm{FWHM}_x \cdot \rm{FWHM}_y}$ (shown as green line). Consequently, we take the observed frequency roll-off of the measured FWHM to define $d_L$ in our measurement system. Best agreement with direct measurements of the FWHM is observed for $d_L$ = 30 mm (calculated from Eq.~(\ref{eqn:Df}), shown as black solid line in \Figrefe{fig:antenna_parameters}). 

In \Figreff{fig:antenna_parameters} we plot the calculated Gaussian beam coupling efficiency $\eta_{gauss} (\nu)$ obtained from the numerical solution of Eq.~(\ref{eqn:effgauss}) for each patch antenna (A1-A4) taking the simulated effective areas $A_{eff}$ and the experimentally determined minimum FWHM into account. Besides that, we compare the solution of Eq.~(\ref{eqn:effgauss}) with values of $\eta_{gauss}$ obtained from the direct evaluation of the numerical area integral for $P_{ant}$ (see Eq.~(\ref{eqn:Pant}) in Appendix A) and $P_{gauss}$ (Eq.~(\ref{eqn:power})) using (i) the measured and (ii) an ideal symmetric Gaussian beam intensity profile $I_{gauss}(x,y)$. The latter is determined from the minimum FWHM (Eq.~(\ref{eqn:Df})) at the respective frequencies. From \Figreff{fig:antenna_parameters}, even for the A3 and A4 with larger effective antenna area, only about 10\% of $P_{in}^{op}$ can be coupled into the antenna element in the best case. 
 Finally, the electrical THz input power at the source transistor terminal $P_{in,t}^{el}$ (see \Figrefa{fig:MOSFET_HDM})) is obtained from \cite{Bauer2019,Ikamas2018,Boppel2012} 
\begin{align}
\label{eqn:etaM}
\begin{split}
P_{in,t}^{el} & = \eta_m(\nu) \cdot P_{in,ant}^{el}  \\ 
              & = \frac{4\cdot \rm{Re}[Z_{ant}]\cdot\rm{Re}[Z_{t}]}{|Z_{ant}+Z_{t}|^2} \cdot P_{in,ant}^{el} \, ,
\end{split}
\end{align}
where $\eta_m(\nu)$ accounts for the impedance mismatch of the antenna impedance $Z_{ant}(\nu)$ (see \Figrefc{fig:antenna_parameters}) to the transistor channel impedance $Z_{t}(\nu)$. Best electrical power coupling to the transistor is ensured whenever conjugate impedance matching conditions are present $Z_{ant} = Z_{t}^*$. Note, that impedance mismatch effects ($\eta_m(\nu)$) are inherently embedded in the circuit simulation software ADS (e.g. by taking a Power Source at a certain frequency with complex impedance) and therefore, for the simulations presented here, we use $P_{in,ant}^{el}$ as THz power input parameter.
Besides the simulated antenna parameters $Z_{ant}(\nu)$, $A_{eff}(\nu)$ as well as the Gaussian beam coupling efficiency $\eta_{gauss}(\nu)$ all other unknown input simulation parameters, in case of ADS-HDM, are obtained from DC drain-to-source resistance ($R_{DS}$) measurements. The measured resistance curves are fitted to a simple serial resistance model for the MOSFET channel, which is based on the UCCM. The extracted simulation parameters are shown in Table~\ref{tab:FETparameter}.\textcolor{black}{\footnote{\textcolor{black}{Note that each detector is connected in series to a different short-circuit protection FET for which we assumed $R_{p,ext}$ = 300 $\Omega$. The exact residual resistance of the protection FET is not known to us and might vary between the devices. This can influence the extracted values for $R_C$.}}}

\section{Simulations and Comparison with Measurements}
\begin{figure}[!t]
\centering
\includegraphics[width=3.5in]{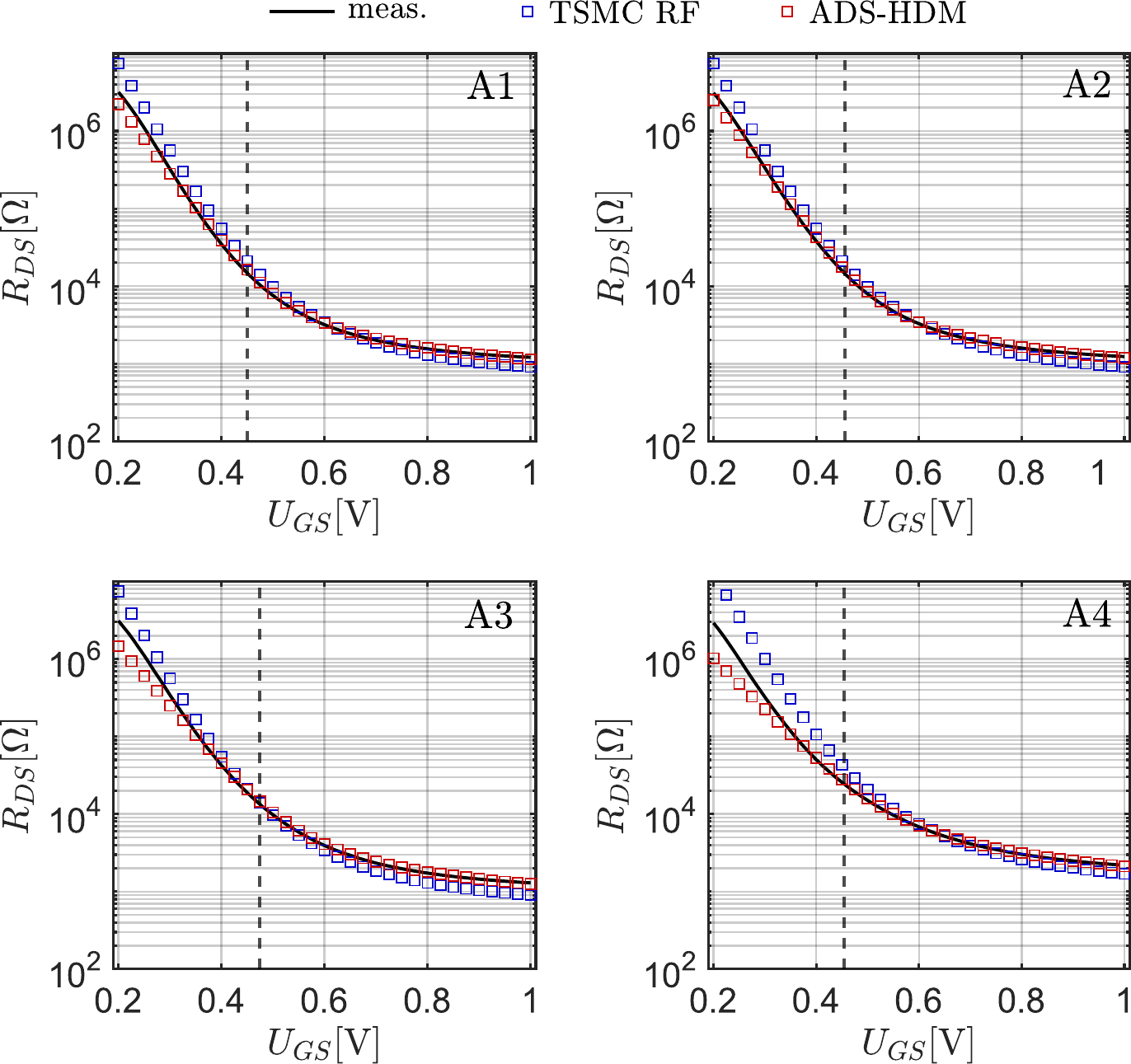}
\caption{Comparison of the measured DC drain-to-source resistance \textcolor{black}{(black line)} with simulations obtained from TSMC RF \textcolor{black}{(blue open squares)} and ADS-HDM \textcolor{black}{(red open squares)} for the different Si CMOS TeraFETs (A1-A4). $U_{DS}$ is set to 10 mV for all measurements and simulations. The dashed vertical line represents the extracted $U_T$ for the respective devices using a fit model based on the UCCM (see Table~\ref{tab:FETparameter}). For both models we assume a fixed residual resistance of $R_{p,ext}$ = 300 $\Omega$ for a short-circuit protection FET, which is connected in series with the drain electrode of all respective transistors.}
\label{fig:DC_resistance}
\end{figure}

\begin{figure}[!t]
\centering
\includegraphics[width=3.5in]{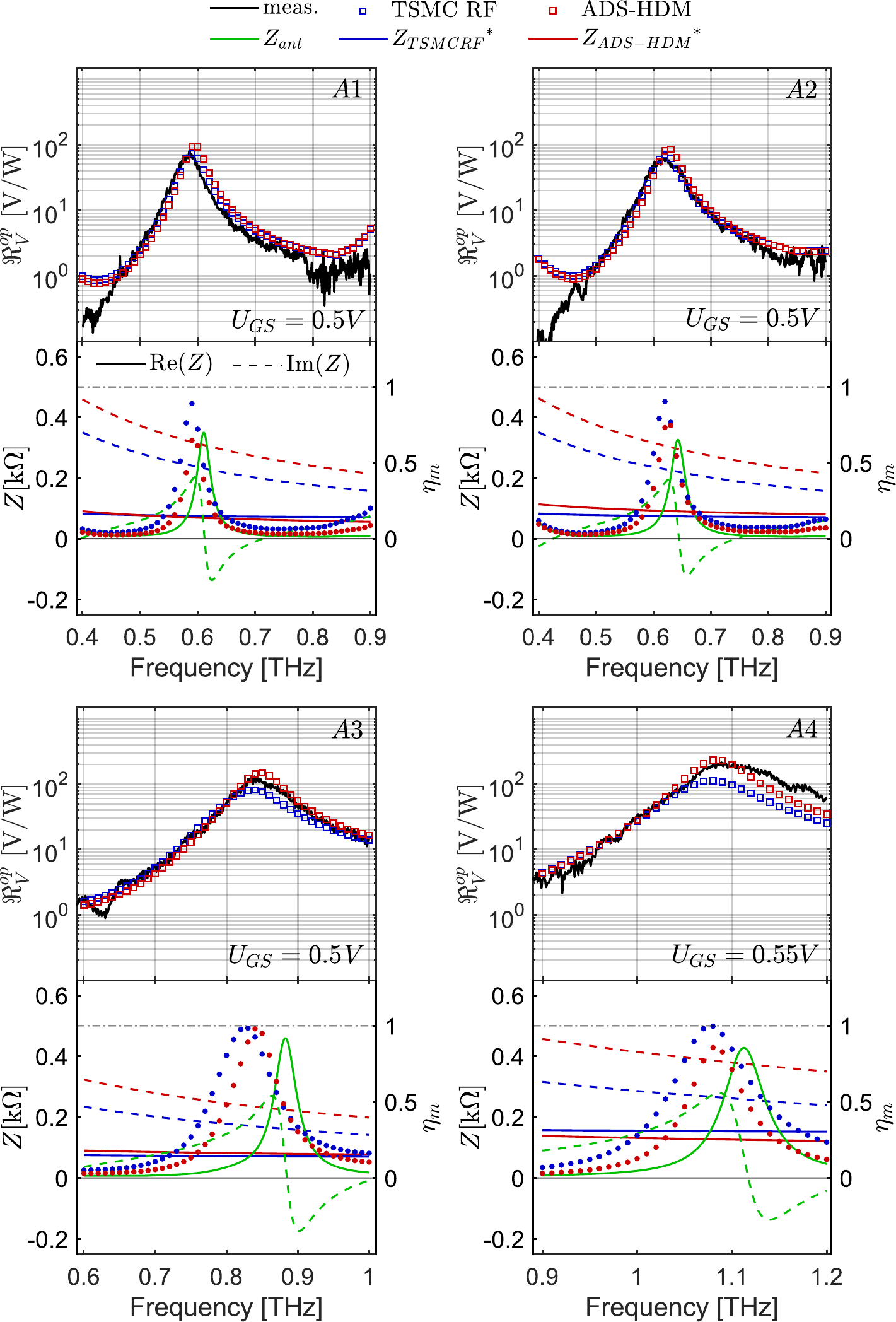}
\caption{Top panels: Comparison of the measured frequency-dependent optical voltage responsivity (black line) with simulations using TSMC RF (blue open squares) and ADS-HDM (red open squares) for devices A1-A4. Bottom panels: Comparison of the simulated frequency-dependent real (solid lines) and imaginary (dashed lines) antenna impedance $Z_{ant}$ (green) with corresponding simulations of the complex conjugate transistor channel impedance ${Z_t}^*$ using TSMC RF (blue) and ADS-HDM (red) for the different Si CMOS TeraFET devices. Estimated impedance mismatch factor $\eta_m$ (dots, right axis) is calculated from Eq.~(\ref{eqn:etaM}) for TSMC RF (blue) and ADS-HDM (red).}
\label{fig:Optical_Responsivity_Impedance}
\end{figure}

In order to validate the implementation technique and the fitting routine discussed previously we compared the simulated $R_{DS}$ as a function of $U_{GS}$ with measurements. In \Figref{fig:DC_resistance}, $R_{DS}$ is shown for TSMC RF and ADS-HDM together with the respective measured data for a fixed drain-to-source voltage of $U_{DS} = 10~$mV. We observe good agreement between the ADS-HDM and data, which is a direct consequence of the specific simulation parameter extraction performed for each device separately (see Table~\ref{tab:FETparameter}). Contrary TSMC RF exhibits only near-quantitative agreement with data close to $U_T$. In the inversion regime ($U_{GS}$\textgreater$U_{T}$) the measured $R_{DS}$ are slightly higher than predicted by TSMC RF. To the best of our knowledge, the discrepancy originates from globalized simulation parameters of the TSMC RF foundry model. These globalized simulation parameters are only adapted depending on the CMOS fabrication process (here CLN65LP, type 1.2V\_mLow\_Vt\_MOS) and the device dimensions (length and width of the channel). Therefore, the foundry model can not account for possible variations (e.g. due to imperfections, defects etc.) of the electrostatic properties within a foundry fabrication run. 
\begin{table}[!t]\centering
  \caption{Extracted simulation parameters for the ADS-HDM of the  different Si CMOS TeraFETs obtained through a fitting model based on the UCCM. $W$ is the width of the channel. $R_C$ is the contact resistance for Source and Drain contact respectively. $\gamma$ is the non-ideality factor.}
\begin{tabular}{rrrrrrr}\hline
Device & $W$ & $\tau_p$ & $U_T$  & $\gamma$  & $R_C$\hspace{5pt} & $\mu$\hspace{15pt} \\
(Patch length)  & ($\mu$m)& (fs) & (V) &  & ($\Omega$)  & ($\frac{\text{cm}^2}{\text{Vs}}$) \\
\hline \hline
A1 (130 $\mu m$)& 1  & 19.14 & 0.45  &  1.9 & 36.37 & 129.48  \\ 
 \hline
A2 (122 $\mu m$)& 1  & 19.67   &  0.456  & 1.89 & 61.25 & 133.09  \\ 
\hline
A3 (84 $\mu m$)& 1  & 18.84  &  0.474  & 2.22 &  63.5 & 127.44  \\ 
\hline
A4 (58 $\mu m$)& 0.45 & 20.38  & 0.454 &  2.56  & 82.46 & 137.85  \\ 
\hline\end{tabular}\label{tab:FETparameter}%y
\end{table}%
A comparison of the frequency-dependent measured optical voltage responsivity (top panels) of devices A1-A4 with the respective simulations -- TSMC RF and ADS-HDM -- can be seen in \Figref{fig:Optical_Responsivity_Impedance}. \textcolor{black}{The final unknown simulation parameter of the ADS-HDM -- the total parasitic capacitance per micron channel width $C_{ov+fr,W^{-1}}$, which includes the overlap $C_{ov}$ and fringe $C_{fr}$ capacitances -- is determined from the comparison of the measured responsivity roll-off of the detector characterized at the highest THz frequencies (A4) with the respective ADS-HDM simulations. Best agreement is found for $C_{ov+fr,W^{-1}} = 0.75\,\rm{fF/\mu m}$. Similar values have been reported for Si CMOS TeraFETs manufactured with the 150-nm technology by TSMC \cite{Boppel2012}. Besides that, in simulations of the detector voltage response $\Delta U_{DS}$, the calculation of the DC read-out is performed against an effective loading impedance $|Z_L| = |(1/R_{in}+i\cdot 2\pi f_{mod} C_L)^{-1}|\approx 440 \, \rm{k\Omega}$ \cite{Sakowicz2011}, taking into account the lock-in amplifier's input impedance $R_{in}$ (10~M$\Omega$) and the effective capacitance of the cable ($C_L=0.5\, \rm{m} \cdot 0.95 \,\rm{pF/m}$).} 

\begin{figure}[!t]
\centering
\includegraphics[width=3.5in]{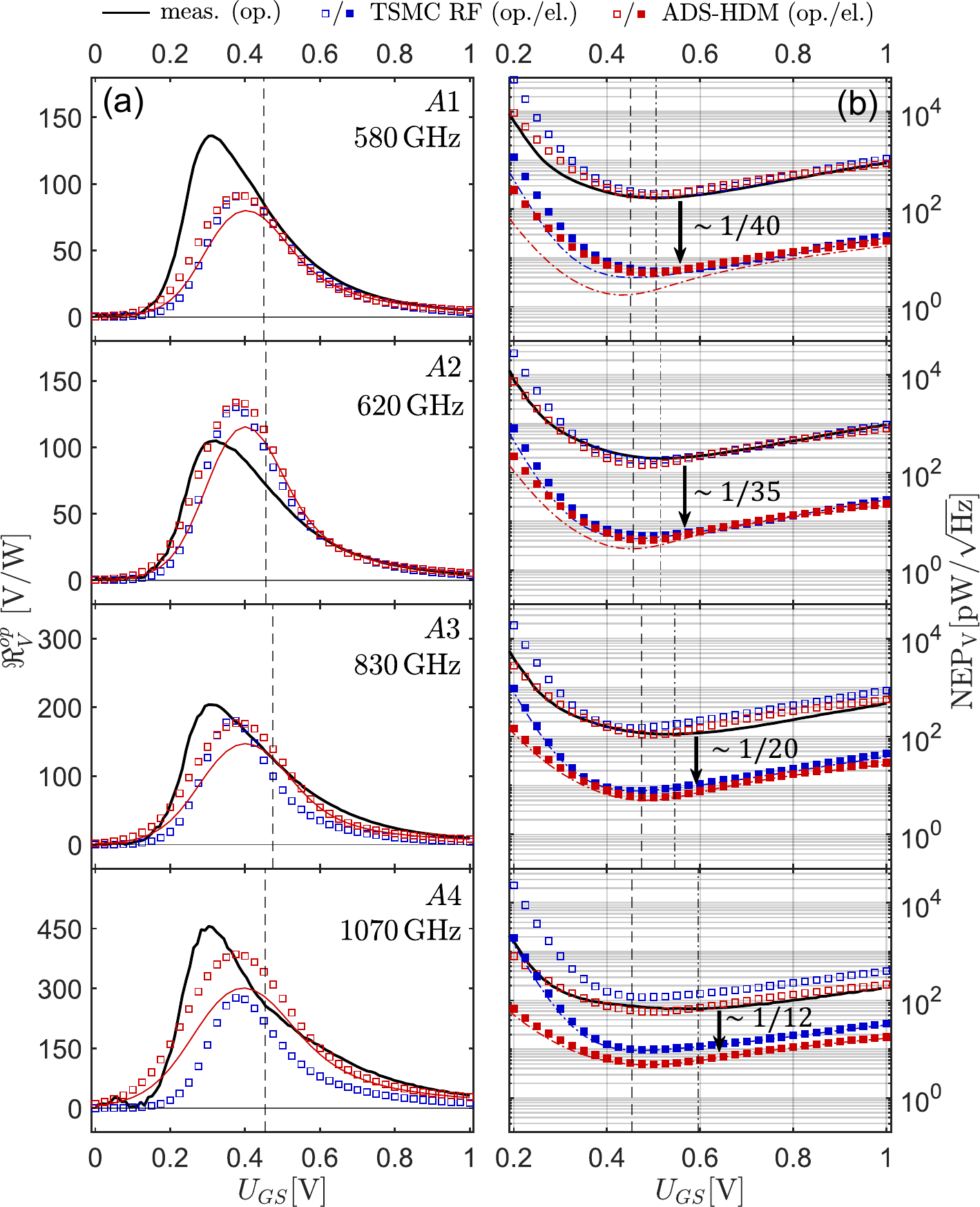}
\caption{Left panels: (a) Comparison of the measured voltage-dependent optical responsivity $\mathcal{R}_{V}^{op}$ (black line) at constant excitation frequency with corresponding simulations obtained from TSMC RF (blue open squares) and ADS-HDM (red open squares) for A1-A4. The vertical dashed line represents the extracted $U_T$ from $R_{DS}$ measurements in case of the ADS-HDM (see Table~\ref{tab:FETparameter}). The pure resistive self-mixing detector response predicted by the ADS-HDM is plotted as red line. Right panels: (b) Comparison of the measured optical voltage-dependent NEP ($\rm{NEP}_{V}^{op}$) (black line) at constant excitation frequency with corresponding simulations using the TSMC RF foundry model (blue open squares) and our in-house tool ADS-HDM (red open squares) for the different Si CMOS TeraFET device. The vertical dashed-dotted line represents minimum $\rm{NEP}_{V}^{op}$ of the respective devices. \textcolor{black}{TSMC RF (blue filled squares) and ADS-HDM (red filled squares) simulations of the voltage-dependent electrical NEP ($\rm{NEP}_{V}^{el}$) with regards to the maximum available power delivered from the antenna $P_{in,ant}^{el}$. In addition, the simulated electrical NEP calculated with regards to the available power at the source transistor terminal $P_{in,t}^{el}$ is depicted for TSMC RF (blue dashed line) and ADS-HDM (red dashed line).}}
\label{fig:Optical_Responsivity_NEP_VGsweep}
\end{figure}

We obtain a near-quantitative agreement between both models and measurements at low THz frequencies (Device A1 and A2). Both models predict magnitude, peak position and the roll-off of the measured signal quantitatively. Only far away from the peak (at 0.4~THz for A1 and A2) we observe a discrepancy between data and simulations. As presented in \Figrefc{fig:antenna_parameters} the EM wave simulations of the antenna impedances $Z_{ant}$ suggest an additional antenna resonance at $\sim0.35$~THz, which is not observed in the THz characterization measurements. At higher THz frequencies (Device A3 and A4) the predicted peak position and magnitude of the optical voltage responsivity of both models starts to differ. The observed divergence can be related mainly to different predicted losses due to imperfect conjugate impedance matching ($Z_{ant} \neq Z_t^*$) between $Z_{ant}$ and $Z_t$. The latter is determined by \textcolor{black}{either} TSMC RF or, in case of ADS-HDM, by the hydrodynamic transport equations (Eq.~(\ref{eqn:HDM})) together with the UCCM (Eq.~(\ref{eqn:UCCM})). In the bottom panels of \Figref{fig:Optical_Responsivity_Impedance} the real part and imaginary part of the simulated complex conjugate transistor channel impedance ${Z_t}^*$ is \textcolor{black}{depicted} together with $Z_{ant}$ for each device. In addition we show the calculated impedance mismatch factor $\eta_m$ (Eq.~(\ref{eqn:etaM}), shown as dots, right axis). Both models suggest that - especially in case of device A3 - perfect conjugate impedance matching ($\eta_m\sim1$) is reached. Remarkably both models predict very similar ${Z_t}$. Despite the fact that TSMC RF has not being adapted specifically for the THz frequency range, it exhibits good agreement with measurements and with ADS-HDM. 
In addition to the frequency-dependent measurements we also measured and simulated the optical voltage responsivity $\mathcal{R}_{V}^{op}$ as a function of gate voltage $U_{GS}$. In \Figrefa{fig:Optical_Responsivity_NEP_VGsweep}, we compare the measured optical voltage responsivity of devices A1 to A4 % (continuous black lines) 
with the results of simulations with ADS-HDM %(red opens squares)
and the TSMC RF %(blue open squares) 
for fixed excitation frequencies. Again, we observe \textcolor{black}{good} agreement between both models and measurements. Best agreement is obtained in the inversion regime above $U_T$ (marked as vertical dashed line) of the respective transistor, while agreement gets worse below $U_T$, \textcolor{black}{where $\mathcal{R}_{V}^{op}(U_{GS})$ peaks.} \textcolor{black}{The deviations are attributed to unknown details of power-coupling and $U_{GS}$-dependent impedance matching.} In addition to results for the full-fledged ADS-HDM, we also plot the predicted purely resistive-mixing response of the transistor, %(shown as continuous red line) 
where the diffusive and plasmonic terms of the HDM are neglected and \textcolor{black}{only the low-frequency limit is considered ($\partial_t j \ll j/\tau_p$), such that} Eq.~(\ref{eqn:Momentum Balance}) is reduced to $j=\sigma \cdot E_x$, with the conductivity $\sigma=(q^2 \tau_p n)/m^* $. The difference between the full ADS-HDM and the purely resistive self-mixing model increases with frequency. \textcolor{black}{The responsivity as predicted by the full model is larger than that of the purely resistive mixing, albeit not by a fixed factor for all values of U$_{GS}$.} The maximal difference reaches values of about 20\% for device A4 at 1070~GHz and $U_{GS}$ = 0.4~V. \textcolor{black}{This increase is evidence that diffusive and plasmonic features cannot be neglected as the frequency rises. Their influence, however, does not become dominant up to the highest frequencies of our studies, and an approximative treatment of rectification by purely resistive mixing yields fair results.} \\
The presented results of measurements and simulations of the optical voltage responsivity have only considered the measured detector response with respect to the available optical THz input power $P_{in}^{op}$. For a detector \textit{sensitivity}, noise contributions to the detected signal must be taken into account. The most important figure of merit in this context is the noise-equivalent power (NEP). The NEP is defined as the ratio of the detectors voltage or current noise spectral density ($V_N$ or $I_N$) in a 1 Hz bandwidth ($\Delta f$) and the detectors voltage or current responsivity, respectively. It can be defined with regards to the optical or electrical detector responsivity ($\mathcal{R}_{V}^{op}$ or $\mathcal{R}_{V}^{el}$). For measurements of the voltage response we have
\begin{align}
\label{eqn:NEP_op_el}
\begin{split}
\rm{NEP}_{V}^{op} & = \frac{V_N}{\mathcal{R}_{V}^{op}} \\ 
\rm{NEP}_{V}^{el} & = \frac{V_N}{\mathcal{R}_{V}^{el}} \, ,
\end{split}
\end{align}
\textcolor{black}{where the voltage noise spectral density $V_N$, in case of zero drain-to-source bias ($U_{DS}=0$~V) during THz detection operation, is dominated by Johnson–Nyquist (thermal) noise, such that $V_N=\sqrt{4 k_B T R_{DS} \Delta f}$~\cite{Lisauskas2013,Bauer2014}. This assumption is based on noise characterization measurements of Si CMOS TeraFETs manufactured by TSMC's 90-nm technology \cite{Bauer2014}}. 
In \Figrefb{fig:Optical_Responsivity_NEP_VGsweep} we present a comparison of the measured optical NEP with simulations obtained by the TSMC RF and the ADS-HDM. Both models correctly predict the overall magnitude and gate voltage dependency of the measured $\rm{NEP}_{V}^{op}$ close and above the $U_T$. Below threshold and at higher THz excitation frequencies (A3 and A4) the ADS-HDM achieves better comparability with data. In order to show possible improvements of the detector performance (e.g. by enhancing the Gaussian beam coupling efficiency $\eta_{gauss}$ or reducing the back-scattering effect of the antenna) we present simulations of the electrical $\rm{NEP}_{V}^{el}$, which represents the intrinsic lower limit of the detector sensitivity. \textcolor{black}{We perform this calculation for two cases, where the electrical responsivity is referred (i) to the available power originating from the antenna element $\mathcal{R}_{V}^{el} = \Delta U_{DS} /P_{in,ant}^{el}$ and (ii) to available power at the Source transistor terminal $\mathcal{R}_{V}^{el} = \Delta U_{DS}/P_{in,t}^{el}$. In the latter impedance mismatch losses $\eta_m$ are excluded. These losses play a more dominant role at lower THz frequencies, where the device impedance $Z_t$ changes more strongly with applied gate voltage.} As already indicated by the estimated values of $\eta_{gauss}$ (see \Figreff{fig:antenna_parameters}) and also visible in \Figrefb{fig:Optical_Responsivity_NEP_VGsweep} there is more room for improvements of the detector sensitivity for A1 and A2 - with respective small effective antennas areas (see \Figrefd{fig:antenna_parameters}) - then for A3 and A4. 
\textcolor{black}{For example, if $\eta_{gauss}$ is increased to $\sim1$ by reducing the FWHM of the Gaussian beam intensity profile in the antenna plane using common methods such as dielectric lenses (placed here on the patch antenna), one can expect that the detector performance increases by a factor of 40 for A1 and 12 for A4.} 
\begin{table}[!h]\centering
  \caption{Measured minimum $\rm{NEP}_{V}^{op}$ in pW/$\rm{\sqrt{Hz}}$ at respective gate voltages $U_{GS}$ for the detectors A1-A4 as well as simulated values obtained from TSMC RF and ADS-HDM for $\rm{NEP}_{V}^{op}$ (op.) and $\rm{NEP}_{V}^{el}$ (el.) with regards to $P_{in,ant}^{el}$.}
\begin{tabular} {|p{1.4cm}|p{1.3cm}|p{1.3cm}|p{1.3cm}|p{1.4cm}|}\hline
Device &A1&A2&A3&A4\\
Frequency &580 GHz&620 GHz& 830 GHz&1070 GHz\\
$U_{GS}$&0.5 V&0.52 V&0.55 V&0.6 V\\
\hline \hline
measured & 169.9 (op.) & 194.7 (op.) & 111 (op.) &  67.17 (op.) \\ 
 \hline
TSMC RF & 208.2 (op.) \newline 5.25 (el.)  & 183.3 (op.) \newline 5.27 (el.)  &  180.6 (op.) \newline  9.3 (el.) & 137.74 (op.) \newline 11.57 (el.) \\ 
\hline
ADS-HDM & 189.9 (op.) \newline 4.93 (el.) & 155 (op.) \newline 4.46 (el.) &  121.5 (op.) \newline 6.25 (el.) & 71.42 (op.) \newline 6 (el.) \\
\hline\end{tabular}\label{tab:NEPvalues}
\end{table}%
In Table~\ref{tab:NEPvalues} the best determined optical NEPs for the respective detectors A1-A4 are shown together with the predictions by the TSMC RF and the ADS-HDM at the respective frequencies and gate voltages. In addition we present the simulated electrical NEP (with regards to the $P_{in,ant}^{el}$) of the respective devices at the same frequency and gate voltage conditions. The simulations suggest that a minimum NEP of $ 5-6 \,\rm{pW}/\rm{\sqrt{Hz}}$ could be achieved at room temperature in ideal optical coupling conditions ($\eta_{op}$~=~1, $\eta_{gauss}$~=~1, $\eta_{scat}$~=~1) and $10-12 \,\rm{pW}/\rm{\sqrt{Hz}}$ if the backscattering effect of the antenna is taken into account ($\eta_{op}$~=~1, $\eta_{gauss}$~=~1, $\eta_{scat}$~=~0.5). This in close agreement to the purely analytical calculations of the electrical NEP for 150-nm Si CMOS TeraFETs as presented in \cite{Boppel2012}. 
\section{Conclusion}
\textcolor{black}{In summary, the detector voltage responsivity and sensitivity of four 65-nm Si CMOS TeraFETs in the THz frequency range of 0.4 and $1.2\,$THz and the gate voltage range of 0 to $1\,$V were investigated in detail using two transistor transport models - the TSMC RF foundry model and our in-house ADS-HDM circuit implementation. The THz characterization measurements were performed under nitrogen atmosphere and the accurate peak-to-peak response of our detectors was determined from time-domain data acquisitions in order to ensure a suitable database for both models. The amount of THz power injected into the transistor terminals by the Gaussian beam via the antenna was calculated based on geometric constraints. A near-quantitative agreement between model and experiment was observed over a wide range of THz excitation frequencies and gate voltages. The simulations indicate that the 65-nm Si CMOS TeraFETs perform close to perfect conjugate impedance matching and further improvement of their detector performance down to $\sim 10 \,\rm{pW}/\rm{\sqrt{Hz}}$ at room temperature can be expected if the efficiency of Gaussian beam coupling is improved. 
By comparing the two models with each other and with experimental I/V and THz characterization, we find that the TSMC RF foundry model is applicable beyond its actual design frequency range for 5G communications and 110 GHz millimeter-wave applications. It can be used with confidence to make predictions about detector performance in the investigated operating range of our Si CMOS TeraFETs up to 1.2~THz. However, the experimental data of A3 and A4 and their comparison with the simulations \textcolor{black}{suggest} that the TSMC RF model may lose its validity \textcolor{black}{at frequencies above 1.2~THz} at which the \textcolor{black}{plasmonic mixing terms in Eq.~(\ref{eqn:Momentum Balance}) \textcolor{black}{are likely to} begin to dominate the device response.}
Our work extends recent advances in compact SPICE circuit modeling of Si MOSFETs in the THz frequency range \cite{Liu2019Spice} by two more simulation approaches. In this work, we compare our simulations directly with experimental data and achieve near-quantitative agreement over a wide operating range. We conclude that both models can be used for modeling and designing MOSFET integrated circuits for operating frequencies of at least 1.2 THz (of special interest for 6G communications, THz imaging, biomedical diagnostics, etc.). Finally, it is worth to mention that the ADS-HDM can be adapted to several key material technologies in the THz research field such as graphene, AlGaN/GaN- (as shown in \cite{Ludwig2019}) and AlGaAs/GaAs HEMTs, as the governing hydrodynamic transport equations only need to be coupled to an charge control model suitable for the respective material technology.}
\appendix
\section{Lock-in amplifier correction factor - Toptica TeraScan 1550}
\begin{figure}[!t]
\centering
\includegraphics[width=3.5in]{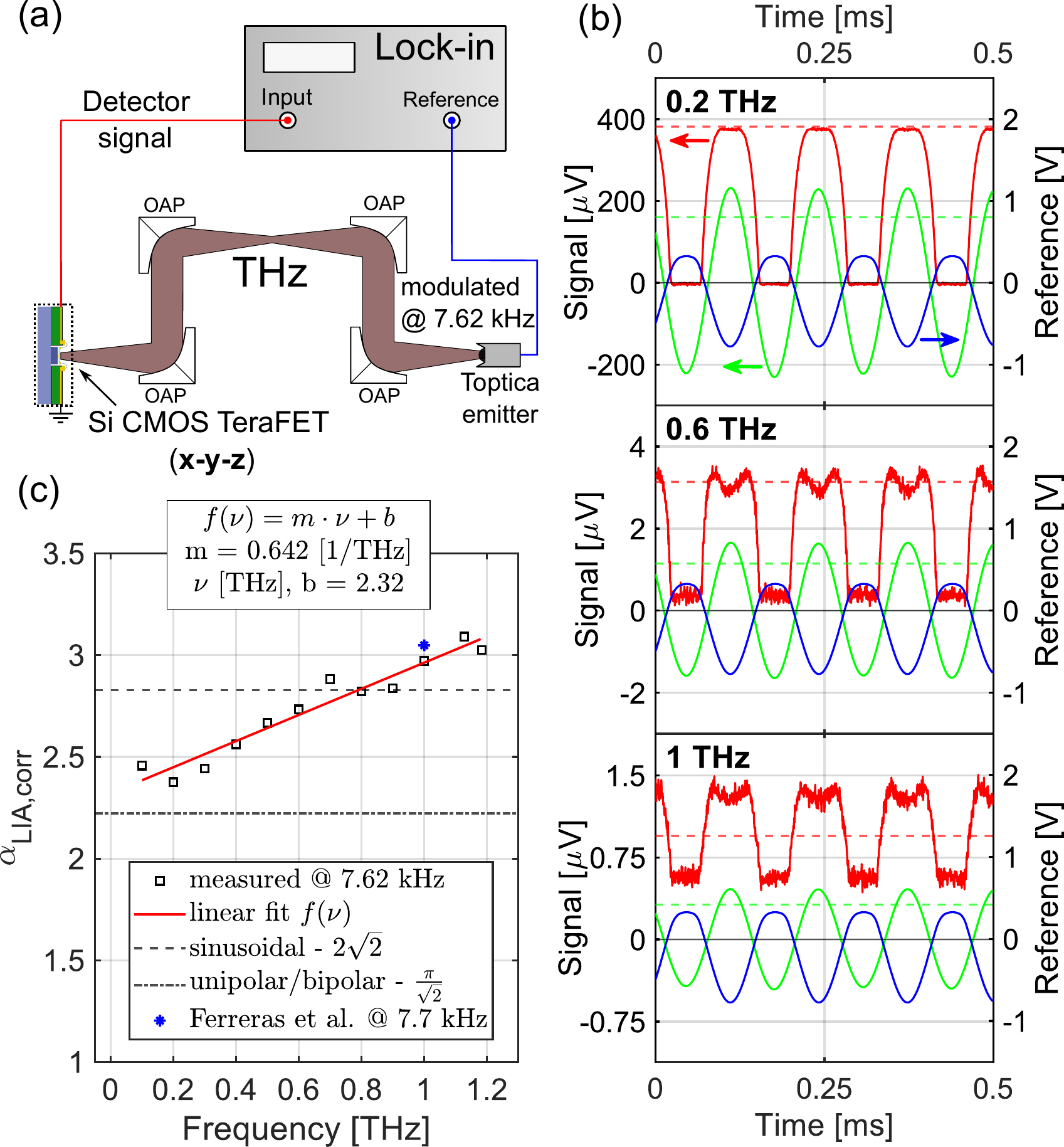}
\caption{(a) Schematic drawing of the measurement setup for optical characterization of the Si CMOS TeraFETs. (b) Fundamental component (green) of the detected waveform (red) extracted from time-domain acquisitions (using an oscilloscope) for a single broadband 65-nm Si CMOS TeraFET at 0.2, 0.6 and 1 THz when illuminating the detector with an electronically chopped (reference shown in blue) InGaAs photomixer (Toptica TeraScan 1550). (c) Extracted lock-in correction factor $\alpha_{LIA,corr}$ as a function of frequency. Calculated values (dashed / dotted lines) for the lock-in correction factor are shown for different types of references/emitter modulations. A reference measurement of $\alpha_{LIA,corr}$ at 1~THz measured with the similar Toptica TeraScan 1550 is shown as blue star\cite{Ferreras2021}.}
\label{fig:LIA_correction}
\end{figure}
In \Figrefa{fig:LIA_correction}, we sketch the measurement setup used to gather the data in this work. The Si CMOS TeraFETs are characterized using the lock-in technique (Ametek DSP 7265) with an electronically modulated InGaAs photomixer emitter (Toptica TeraScan 1550) @ $f_{mod}$~=~7.62 kHz (indicated as blue line). Alignment of the detectors was performed in $xyz$ to guarantee the correct position in the focal point of the final OAP mirror (\textcolor{black}{diameter:} 2''). 
%We note that in principle $\alpha_{LIA,corr}$ should be constant and in agreement with its mathematical prediction when the electronic modulation of the THz emitter works as intended. 
From time-domain acquisitions of the detector signal (red line), we determined the lock-in correction factor $\alpha_{LIA}$ via
\begin{align}
\label{eqn:A_correction_factor}
\alpha_{LIA,corr} & = \frac{\rm{Pk2Pk}(\Delta U_{DS})}{\rm{RMS}(U_{lock-in,1f_{mod}})}  
\end{align}
as a function of the radiation frequency using a single broadband 65-nm Si CMOS TeraFET (log-spiral antenna). \\ In \Figrefb{fig:LIA_correction}, the measured detector signal $U_{DS}$ (red line, left axis), the modulation waveform for the Toptica emitter (blue line, right axis) -- which serves as the reference for the lock-in amplifier --, and the fundamental component of the measured detector signal  $U_{lock-in,1f_{mod}}$ (green line, left axis) -- extracted using Fourier analysis -- are presented for three different radiation frequencies (0.2, 0.6 and 1~THz). In addition, the extracted peak-to-peak value (red dashed line), relevant for the detector calibration and sensitivity measurements, as well as the obtained RMS value of the fundamental component (green dashed line), which is measured by the lock-in amplifier after bandpass filtering at $f_{mod}$~=~7.62 kHz, are shown. 
In \Figrefc{fig:LIA_correction}, we present $\alpha_{LIA,corr}$ as a function of the radiation frequency. A linear fit is used to determine the observed frequency trend, which is relevant for our simulations and correct predictions of $\Delta U_{DS}$. 
\textcolor{black}{We speculate that the unexpected frequency dependence of $\alpha_{LIA,corr}$ is a consequence of frequency-dependent non-linear mixing properties of the Toptica InGaAs photomixer emitter. They manifest, when the photomixer is illuminated by the fiber-coupled dual-color optical beam at different beat-note frequencies $\Delta\nu$ (0.1-1.2~THz). This leads to the observed non-ideal (here non-sinusoidal) modulation of the emitted THz output power and to the change in $\alpha_{LIA,corr}$ with the beat-note frequency. Further investigations, also at higher frequencies, are needed to clarify the origin of the observed effect.}
\section{Gaussian beam coupling efficiency}
\begin{figure}[!ht]
\centering
\includegraphics[width=1.8in]{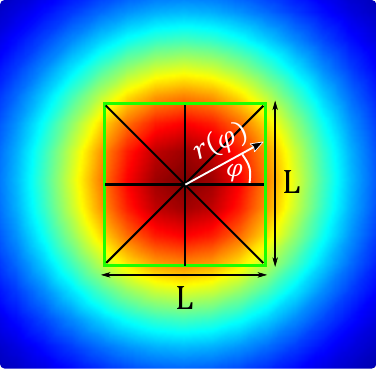}
\caption{Sketch for visualizing mathematical steps for the calculation of the Gaussian beam coupling efficiency. The dimensions of the effective area $ A_{eff} = L^2$ are indicated as a green square.}
\label{fig:Gaussian_beam_eff}
\end{figure}

In order to calculate the total power coupled into the effective area (indicated as green square in \Figref{fig:Gaussian_beam_eff}) of an arbitrary planar antenna we assume a point symmetry of $I_{gauss}(x,y)$ to split the calculation into 8 equal triangles as indicated in \Figref{fig:Gaussian_beam_eff} (black triangles). Together with the parameterization $r(\varphi) = \frac{L}{2}\cdot \cos{\varphi}$ we obtain
\begin{align}
\label{eqn:Pant}
\begin{split}
P_{ant} &= \int\int_{A_{eff}} I_{gauss}(x,y) \,\rm{d}x \rm{d}y  \\
        &= 8 \cdot \int\int_{A_{\Delta}} I_{gauss}(x,y) \,\rm{d}x \rm{d}y \\
        &= 8 I_0 \cdot \int_{0}^{\frac{\pi}{4}} \int_{0}^{r(\varphi)} r \cdot \exp{(-\alpha r^2)} \, \rm{d}\varphi \rm{d}r \\
        &= \frac{4I_0}{\alpha} \cdot \int_{0}^{\frac{\pi}{4}} [1-\exp{(\frac{\alpha \cdot L^2}{4}\frac{1}{\cos{\varphi}^2})}] \,\rm{d}\varphi \\
        &= \frac{I_0}{\alpha} \cdot [\pi-4\beta]. 
\end{split}
\end{align}
With the definition of the Gaussian beam coupling efficiency $\eta_{gauss} = \frac{ P_{ant}}{P_{gauss}}$ we finally derive the equation set Eq.~(\ref{eqn:effgauss}).
%\section{}
\section*{Acknowledgment}
This work was funded by DFG RO 770/40-1, 770/40-2 and 770/43-1.


\bibliography{THz}
\bibliographystyle{unsrt}

\end{document}